\documentclass[10pt,twocolumn,letterpaper]{article}
\usepackage{multirow} 
\usepackage{natbib}
\usepackage[review]{iccv}      

\definecolor{iccvblue}{rgb}{0.21,0.49,0.74}
\usepackage[pagebackref,breaklinks,colorlinks,allcolors=iccvblue]{hyperref}

\def\paperID{9607} 
\def\confName{ICCV}

\title{RoSMM: A Robust and Secure Multi-Modal Watermarking Framework for Diffusion Models}

\author{ZhongLi Fang\\
Fudan University, Shanghai, China\\
{\tt\small zlfang22@m.fudan.edu.cn}
\and
*Yu Xie\\
The Purple Mountain Laboratories, Nanjing, China\\
\and 
*Ping Chen\\
Fudan University, Shanghai, China\\
}

\begin{document}
\maketitle
\begin{abstract}

Current image watermarking technologies are predominantly categorized into text watermarking techniques and image steganography; however, few methods can simultaneously handle text and image-based watermark data, which limits their applicability in complex digital environments. This paper introduces an innovative multi-modal watermarking approach, drawing on the concept of vector discretization in encoder-based vector quantization. By constructing adjacency matrices, the proposed method enables the transformation of  text watermarks into robust image-based representations, providing a novel multi-modal watermarking paradigm for image generation applications. Additionally, this study presents a newly designed image restoration module to mitigate image degradation caused by transmission losses and various noise interferences, thereby ensuring the reliability and integrity of the watermark. Experimental results validate the robustness of the method under multiple noise attacks, providing a secure, scalable, and efficient solution for digital image copyright protection.
\end{abstract}

\section{Introduction}
\label{sec:intro}
With the rapid rise of AIGC (Artificial Intelligence-Generated Content) \cite{rombach2022high}\cite{wallace2023edict}, countless synthetic images are now shared across media and the internet. While advanced models like diffusion \cite{nichol2021improved}\cite{watson2022learning}\cite{pandey2022diffusevae} meet the demand for high-quality content, their misuse has raised serious issues around data privacy, copyright protection, and social trust. New watermarking technologies \cite{xiong2023flexible}\cite{fernandez2023stable}\cite{asnani2024promark} provide a simple and effective way to resolve copyright disputes and track information.

\begin{figure}[htbp]
\centering
\includegraphics[width=\columnwidth]{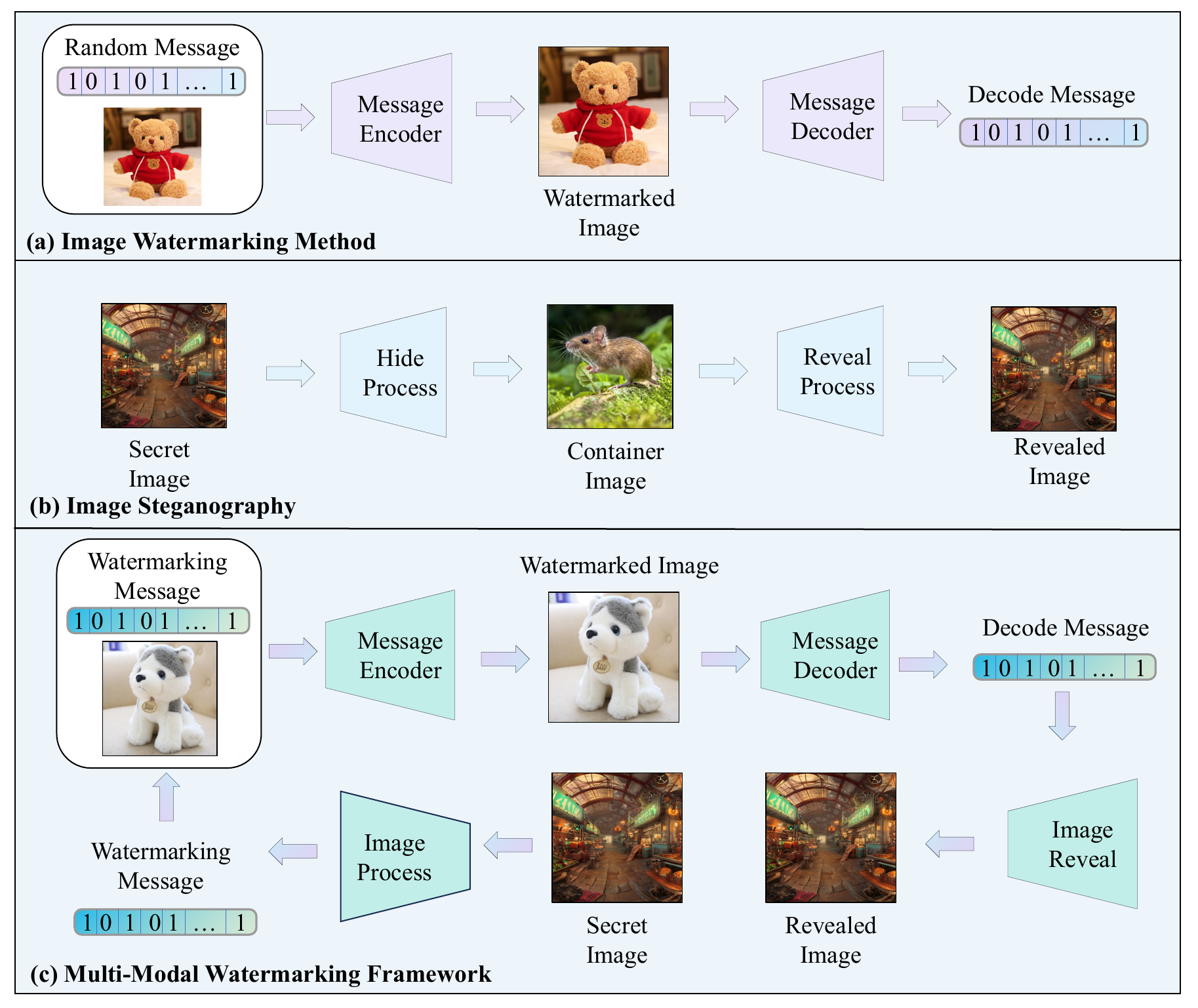}
\caption{Current watermarking methods fall into two main categories: text-based watermarking for traceability and image steganography for embedding information. Our method goes beyond these by converting binary watermarks into more robust image-based watermarks, offering a new multi-modal approach for watermarking in image generation.}
\label{fig:first}
\end{figure}


Watermarking offers a direct method for tracing image information, as it can be embedded explicitly or implicitly within generated images, providing a foundational solution for subsequent copyright protection and information traceability. Existing image watermarking techniques are generally divided into two categories: text-based watermark tracing methods \cite{wen2024tree}\cite{ci2024wmadapter} and image-based steganography \cite{lu2021large}\cite{jing2021hinet}, as illustrated in Figure \ref{fig:first}. Image-text watermarking methods are limited in the amount of information they can convey, typically constrained to 16 to 256 bits, with recent advances \cite{yang2024gaussian} achieving stable transmission of up to 512 bits, approximately 64 characters. In contrast, image steganography embeds secret images within generated images and typically requires noise-free conditions for training, making it vulnerable to real-world disturbances such as noise and non-linear transformations. These limitations hinder the practicality and applicability of image steganography in robust environments \cite{jaini2019sum}\cite{papamakarios2017masked}.


From the above analysis, it is evident that each of these methods has distinct limitations: traditional image-text watermarking, while widely applicable, transmits only a limited amount of information, whereas image steganography can embed a complete secret image, but it involves significant computational costs and lacks robustness against external attacks. In general, existing methods struggle to balance security, robustness, and information density in watermarks The natural question arises: is it possible to combine the security and robustness of text content watermarking with the informational density characteristic of image-based steganography? 

\begin{figure}[htbp]
\centering
\includegraphics[width=\columnwidth]{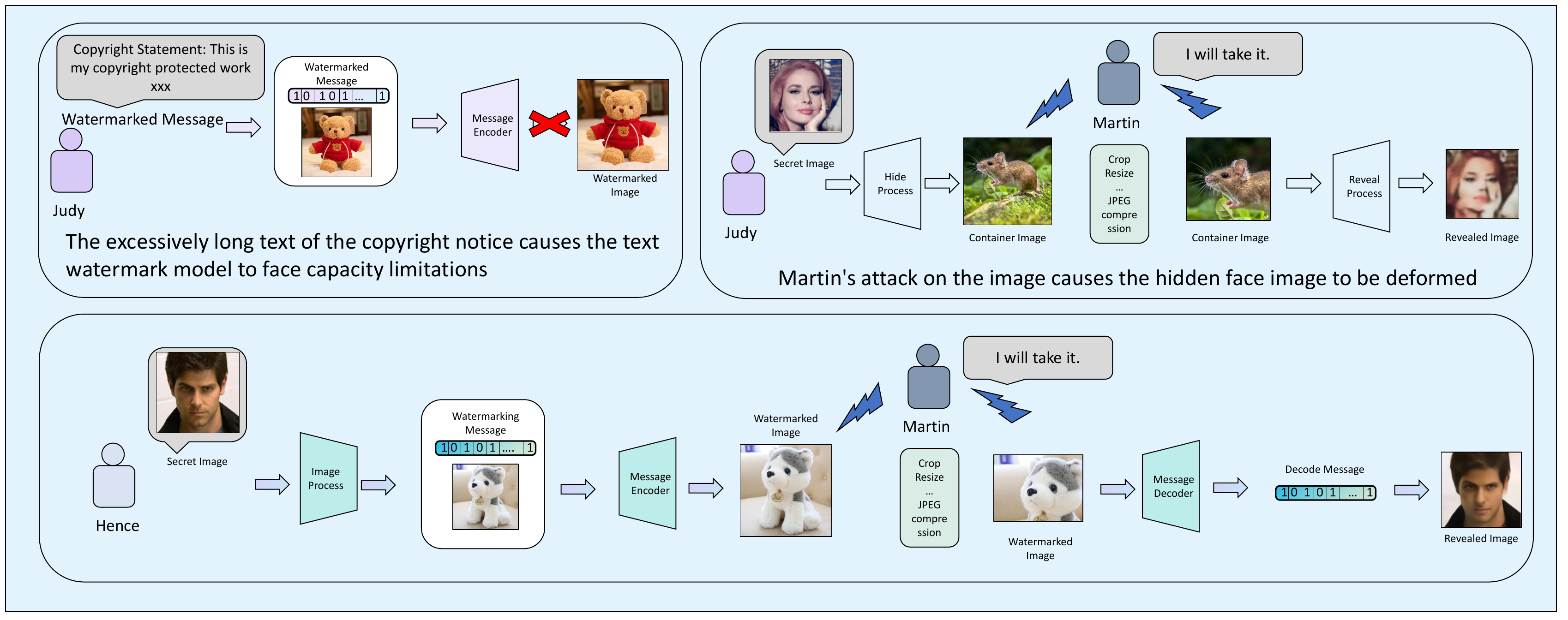}
\caption{Applications of multi-modal watermarking. The proposed method demonstrates exceptional performance in watermark traceability, copyright protection, and secure data transmission. }
\label{fig:second}
\end{figure}

The answer is affirmative. For the concept of vector discretization coding from vector quantized variational autoencoders, this paper introduces an innovative Robust and Secure Multi-Modal Watermarking framework (RoSMM) that not only enables conversion between image and text watermarks but also integrates the advantages of image steganography and text content watermarking, establishing a novel approach to multi-modal watermarking content. Additionally, to further enhance the robustness and readability of the watermark, we designed an image restoration module that performs secondary restoration processing on the watermark information, significantly enhancing the anti-interference capability of the watermark content. Overall, by combining text-based watermarking and image steganography, this framework expands the functionality and scope of watermarking techniques, providing new possibilities for watermark traceability and application versatility.

\begin{figure*}[htbp] 
\centering
\includegraphics[width=\textwidth]{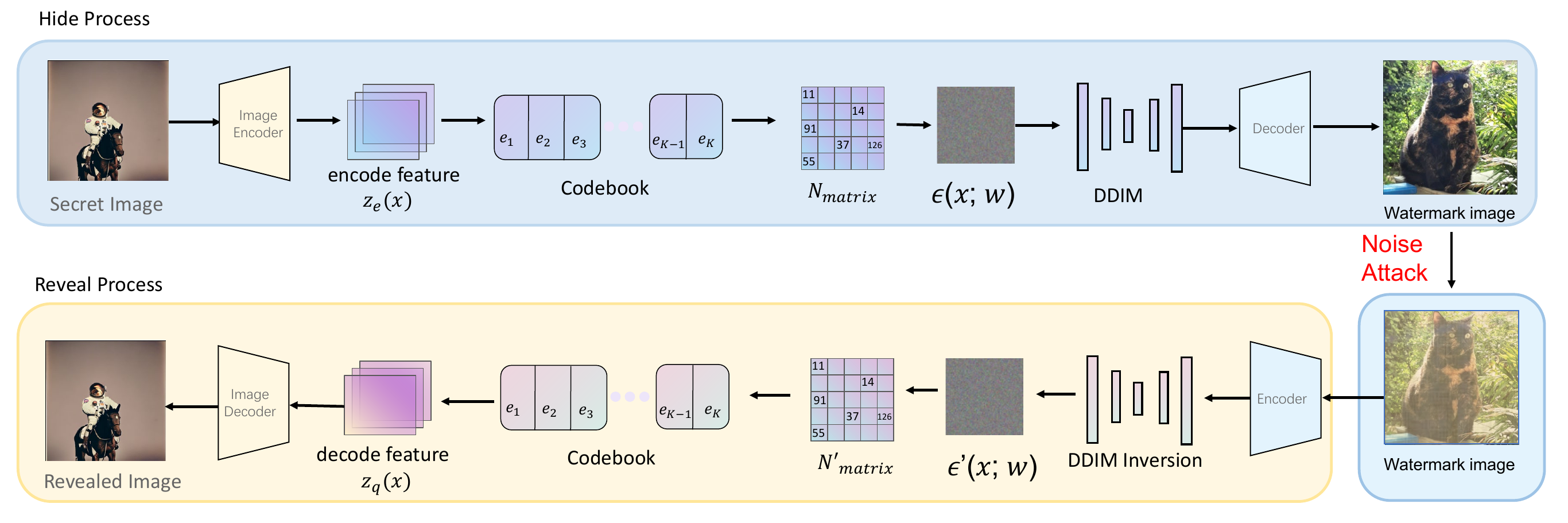} 
\caption{This figure describes in detail the overall framework of the multi-modal watermarking content method proposed in this paper. First, the input image \( x \) is converted into a high-dimensional latent feature representation \( z_e(x) \) through a multi-layer convolutional encoder.
After that, these features are converted into a binary watermark by forming an adjacency matrix \( N_{\text{matrix}} \) through codebook mapping and quantization, and then embedded into the latent features through a diffusion model.
Through the restoration and decoding process, the model finally generates an image containing the watermark information, ensuring the concealment of the watermark and its robustness to attacks.}
\label{fig:third}
\end{figure*}

To the best of our knowledge, we are the first to address and implement the challenging task of multi-modal watermarking content. Moreover, this method can be integrated as a simple add-on module into existing text watermarking technologies, seamlessly transforming them into multimodal watermarking content systems without compromising their original usability and user experience. Our contributions can be summarized as follows:

\begin{itemize}
\item This paper proposes an innovative multimodal watermarking framework that enables seamless conversion between text and image watermark content, effectively bridging the gap between text watermarking and image steganography technologies for the first time.

\item The framework incorporates an innovative image feature restoration module, which significantly enhances the robustness of watermark content during transmission and ensures the integrity of its conversion into image content, even under various noise and interference conditions

\item The multimodal watermarking module can be seamlessly integrated into existing text watermarking technologies, transforming them into multimodal watermark content models without compromising the original user experience or functionality.

\item Comprehensive experiments focusing on the robustness of the watermark demonstrate the superiority and stability of the proposed method compared to existing approaches.
\end{itemize}

\section{Related Work}
\label{sec:related work}
\subsection{Diffusion Models}

Diffusion models \cite{nichol2021improved}\cite{song2019generative}\cite{song2020score} are a class of emerging deep generative models that have demonstrated groundbreaking performance in various applications, such as image synthesis \cite{dhariwal2021diffusion}\cite{saharia2022photorealistic}\cite{ramesh2022hierarchical}, video generation \cite{ho2022imagen}\cite{ho2022video}, and image restoration \cite{saharia2022image}\cite{wang2022zero}\cite{kawar2022denoising}. The core principle of these models is to simulate the physical diffusion process: noise is incrementally added to the data, transforming it into a noisy state, which can then be reconstructed back to its original form through a reverse process. Initially, Ho et al. \cite{ho2020denoising} introduced the denoising diffusion probabilistic model (DDPM), which laid the groundwork for subsequent research \cite{lyu2019advances}\cite{gu2022vector}\cite{lu2022dpm}. Later, Rombach et al. \cite{rombach2022high} proposed latent diffusion models, significantly reducing computational complexity and improving efficiency, thus further advancing the development of diffusion models \cite{kim2022diffusionclip}\cite{meng2021sdedit}\cite{tang2023emergent}. This progress has spurred a wave of innovation within the diffusion model community, leading to many new approaches. \cite{gandikota2023erasing}\cite{wang2024patch} In this paper, our proposed framework also leverages diffusion-based watermarking techniques, harnessing their inherent robustness and security benefits to strengthen our multi-modal watermarking method.

\subsection{Image Watermarking}
Image-text watermarking technology \cite{ahmadi2020redmark}\cite{liu2024mirror}, as a means of embedding textual information within images, is primarily used to insert copyright details, authentication data, or other sensitive text for tracking, verification, and copyright protection purposes. 
Huang et al. \cite{huang2019enhancing} proposed an adaptive image watermarking method that can be directly applied to the outputs of diffusion models. Zhu et al. \cite{Zhu2018HiDDeNHD} introduced a watermark embedding approach based on adversarial training. these methods effectively enhance the robustness and imperceptibility of the watermark, but like other post-processing techniques, it can still significantly affect image quality. Recent approaches, such as the Stable Signature method \cite{fernandez2023stable} and the method proposed by Zhao et al. \cite{zhao2023recipe}, attempt to embed watermarks during the diffusion generation process to mitigate impacts on image quality. However, these methods only allow the extraction of fixed watermark information. Other innovative studies, such as the Tree-ring method \cite{wen2023tree}, aim to embed copyright information by modifying latent representations, thereby eliminating training costs; however, this approach only achieves copyright protection without enabling the transmission of meaningful information. Even the latest models, such as Gaussian Shading \cite{yang2024gaussian}, are limited to transmitting approximately 512 bits of effective information. Consequently, the bit limitations of image-text watermarks restrict the density of content they can carry.

\subsection{Steganography Methods}

Image steganography is a technique for embedding secret information into a host image, aiming to create an information carrier that is difficult to detect. For instance, Zhu et al. \cite{Zhu2018HiDDeNHD} and Zhang et al. \cite{zhang2019steganogan} attempted to introduce adversarial learning methods, utilizing an encode-decode architecture to automatically learn the embedding and recovery of information, in order to explore the balance between image quality and image steganography. In references \cite{lu2021large} and \cite{jing2021hinet}, researchers proposed an innovative design for steganographic models by constructing them as invertible neural networks (INN) \cite{dinh2014nice}\cite{dinh2016density}. Building on this, coverless steganography is an emerging approach that does not modify the host object, making it more challenging for attackers to detect hidden data. Examples of this approach include the techniques found in \cite{luo2020coverless} and \cite{zou2022robust}, which have emerged as notable methods within the field. Currently, researchers are exploring steganography methods that do not rely on host images, aiming to achieve even higher security in communication. However, most existing methods still depend on embedding data into carrier images, which means that perturbation attacks on these images can significantly impact the hidden content.

\section{Methods}
\label{sec:rmethods}

\subsection{Application Scenarios}
As show in Figure \ref{fig:second}, the proposed multi-modal watermarking content method has been validated for its effectiveness across multiple application scenarios. The scenarios involve three models: Image-Watermark Model A, Image Steganography Model B, and the proposed Multi-modal Watermarking Model C. Users Judy and Hence, along with an attacker Martin, tested these models. Judy attempted to embed a copyright mark using Models A and B but found them ineffective under attack. In contrast, Hence successfully embedded a robust watermark using Model C. Even after the attack, the watermark remained intact, proving ownership and demonstrating the robustness of the multi-modal approach.

\subsection{Seamless Text-Image Watermarking}
This section introduces an innovative multi-modal watermarking technique aimed at enabling seamless conversion between text and image watermarks. This method first constructs a codebook to vectorize the image to be embedded, then applies vector quantization to map the image features onto a predefined codebook. Next, an adjacency matrix is generated based on the similarity of discrete vectors, encoding the structural information of the image as binary data to achieve seamless conversion between text and image. Section 3.2.1 presents a mathematical proof of the vector discretization encoding, and introduce its specific application within this method in subsequent chapters.
\subsubsection{Foundations of Discrete Encoding}

\textbf{Encoding}: For an input image, the model extracts key information via a deep neural network, transforming it into a continuous latent representation.

\textbf{CodeBook}:A codebook is a finite set of vectors (codewords) representing the discrete latent space of the input. These pre-learned vectors act as a "vocabulary," mapping image features to discrete codebook vectors during vector quantization, transitioning from continuous to discrete latent space.

\textbf{Decoding}: The selected codebook vector reconstructs or represents the original input.

The mathematical expression of vector quantization can be defined as follows:

\textbf{Encoder Output}: Let \( z_e(x) \) be the encoder’s continuous latent representation for input \( x \).

\textbf{Vector Quantization Layer}: This layer maps \( z_e(x) \) to a discrete latent space by finding the closest vector \( e_k \) in the codebook \( \{e_1, e_2, ..., e_K\} \), where \( K \) is the codebook size:
\begin{equation}\label{eq:1}
z_q(x) = \text{argmin}_k ||z_e(x) - e_k||_2^2
\end{equation}
   where \( ||\cdot||_2 \) denotes the L2 norm (Euclidean distance).

\textbf{Embedding Loss}: The embedding loss \( L \) combines the distance between the encoder output and the quantized vector, and the distance between the quantized vector (via embedding network \( g \)) and the encoder output:
\begin{align}\label{eq:2}
   L = \mathbb{E}_{x,z_e(x),z_q(x)}[\|z_e(x) - z_q(x)\|_2^2] \\
   + \beta \mathbb{E}_{x,z_q(x)}[\|z_e(x) - g(z_q(x))\|_2^2] \nonumber 
\end{align}
Here, \( g \) is the embedding network, and \( \beta \) balances the two terms. The first term is the vector quantization loss (\( L_{\text{embedding}} \)), while the second is the commitment loss (\( L_{\text{commitment}} \)), ensuring the encoder commits to the selected codebook vector and stabilizes training. Together, these terms maintain discrete encoder outputs while optimizing image quality.

\begin{figure}[htbp] 
\centering
\includegraphics[width=\columnwidth]{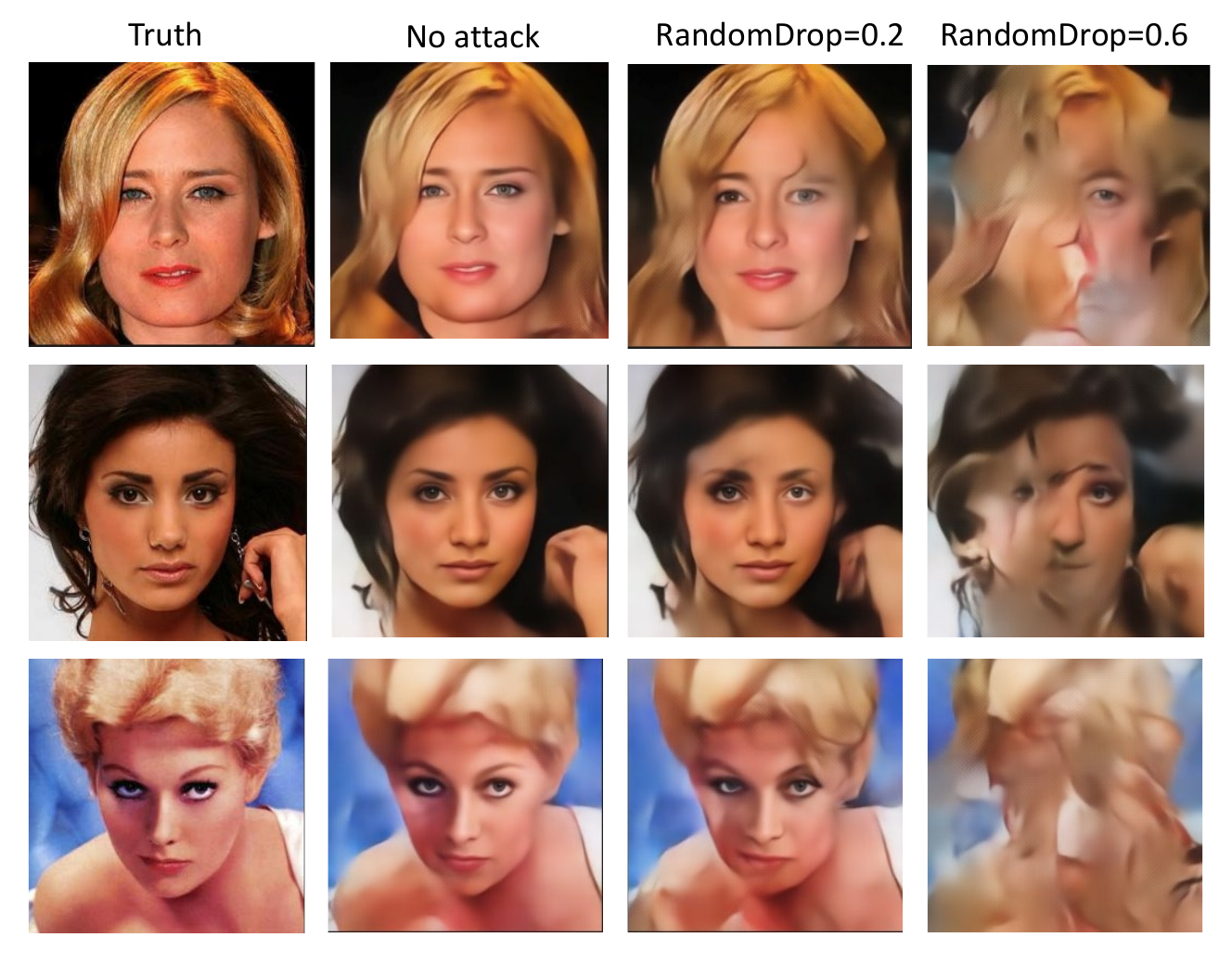} 
\caption{The image watermark imaging effect of the multi-modal watermarking method under attack conditions without image restoration module.}
\label{fig:four}
\end{figure}

\begin{figure*}[htbp] 
\centering
\includegraphics[width=\textwidth]{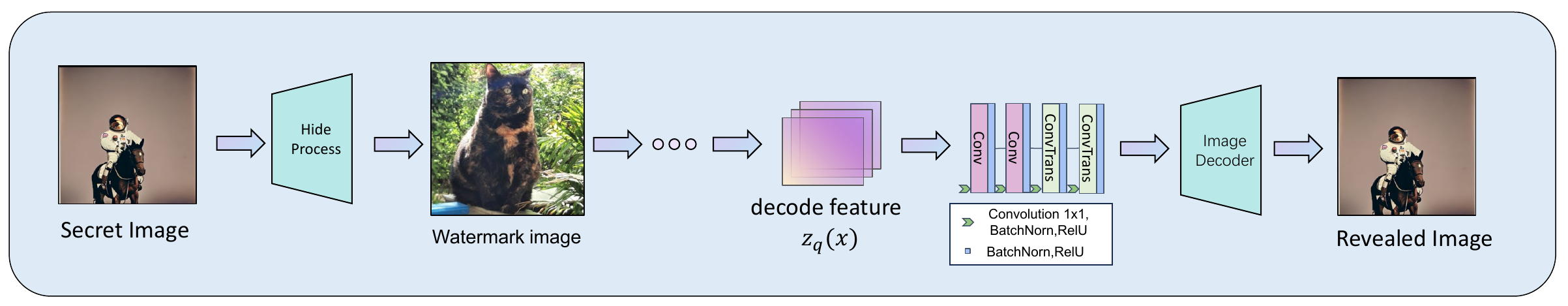} 
\caption{The architecture of the image restoration module. This module removes lossy stain blocks by compressing features and restores image features through upsampling technology, thereby effectively ensuring the imaging effect of the image.}
\label{fig:five}
\end{figure*}


\subsection{Multi-Modal Watermarking Method}

The proposed method utilizes codebook and feature discretization techniques to achieve text-to-image pair conversion. Considering the training overhead and watermark capacity, we selected the Gaussian Shading model as the foundational model within the proposed multi-modal watermarking framework. In summary, the multi-modal watermark framework of this paper is shown in Figure \ref{fig:third}.

During watermark embedding, the input image \( x \) is first transformed into high-dimensional latent features \( z_e(x) \) using a multi-layer convolutional encoder. These features are then quantized into discrete codewords via a codebook, forming an adjacency matrix \( N_{\text{matrix}} \). The matrix is constructed by computing pairwise cosine similarities between latent feature vectors:
\begin{equation}\label{eq:3}
N_{\text{matrix}}(i, j) = \frac{z_e(x)_i \cdot z_e(x)_j}{||z_e(x)_i||_2 ||z_e(x)_j||_2} \quad 
\end{equation}
where \( z_e(x)_i \) and \( z_e(x)_j \) are latent feature vectors, and \( N_{\text{matrix}}(i, j) \) represents the element at the \( i \)-th row and \( j \)-th column of the adjacency matrix.

Next, the adjacency matrix \( N_{\text{matrix}} \) is converted into a binary secret watermark. First, the matrix is resized to a 16×16 dimension to match the required binary watermark length, considering the capacity of the text watermark. Then, each element of the adjacency matrix is transformed into an 8-bit binary representation, and these binary strings are concatenated to form the final binary watermark. This process can be expressed as:
\begin{equation}\label{eq:4}
 w = \text{Concatenate}(\text{Binary}(N_{\text{matrix}}(i, j), 8)) \quad 
\end{equation}
where \( \text{Binary}(N_{\text{matrix}}(i, j), 8) \) denotes the conversion of each adjacency matrix element into an 8-bit binary string, and \( \text{Concatenate} \) signifies the joining of these binary strings to form the binary watermark \( w \).

Subsequently, the binary watermark information is input into a diffusion model, projecting the watermark information onto the latent feature representation using Gaussian shading technology. This process adheres to the standard Gaussian distribution, ensuring the stealth of the watermark, which can be represented as:
\begin{equation}\label{eq:5}
z_w = z_e(x) + \eta \quad 
\end{equation}
where \( \eta \) is Gaussian noise, and \( z_w \) is the latent feature representation containing the watermark information.

Finally, through the restoration module of the diffusion model, the latent feature representation containing the watermark information is converted back into the image space, ultimately generating a watermarked image through the decoder. The decoder utilizes its complex network structure, including transposed convolutional layers and activation functions, to map the latent features back into the original image space.

\begin{table*}[htbp]
\centering
\begin{tabular}{c|c|ccc|ccc|ccc}
\toprule
\multirow{2}{*}{methods} 
 & \multirow{2}{*}{clean}  & \multicolumn{3}{c}{Grussian noise} & \multicolumn{3}{c}{Brightness}  & \multicolumn{3}{c}{Random Crop} \\
\cmidrule(lr){3-5} \cmidrule(lr){6-8}  \cmidrule(lr){9-11}
 & & $\theta$=0.05 & $\theta$=0.1 & $\theta$=0.2 & $\theta$=1 & $\theta$=2 & $\theta$=4 &  $\theta$=0.2 & $\theta$=0.4 & $\theta$=0.8 \\
\midrule
Baluja  (256) \cite{baluja2019hiding} & 28.91 &9.52 &9.48 &9.32 &8.90 &7.30 &6.00 &9.33 &9.21 &9.00 \\
Baluja  (512) \cite{baluja2019hiding} &28.79  &9.54 &9.50 &9.33 &9.55 &7.31 &6.00 &9.27 &9.08 &8.72 \\
HiNet  (256) \cite{jing2021hinet} & 33.37 & 7.81 & 6.51 & 6.13  & \textbf{23.73} & 18.49 & 16.74 & 10.78 & 11.13 & 12.73 \\
HiNet (512) \cite{jing2021hinet} & 33.93 & 6.90 & 5.49 & 5.01   & 21.71 & 18.56 & 15.86 & 10.72 & 11.23 & 14.28  \\
CRoSS (256) \cite{yu2024cross}& 15.62 & 11.50 & 10.75 & 9.61 & 13.09 & 11.11 & 10.59  & 6.43 & 6.84 & 9.59\\
CRoSS (512)\cite{yu2024cross}& 25.30 & 17.04 & 14.57 & 11.61  & 17.70 &14.69 & 12.54  & 6.32 & 6.97 & 10.61 \\
\midrule
RoSMM-w & 22.21 & 12.00 & 10.14 & 9.38 & 20.03 & 19.48  & 17.04 & 9.24 & 9.33 & 13.75 \\
RoSMM(ours) & 20.35 & \textbf{17.32} & \textbf{15.23} & \textbf{12.28} & 20.70 & \textbf{20.69} & \textbf{19.90} & \textbf{10.92} & \textbf{11.23} & \textbf{18.69} \\
\bottomrule
\end{tabular}
\caption{The PSNR (dB) performance results of different methods under various types of attacks. The higher the PSNR value, the better the image quality.}
\label{tab:performance}
\end{table*}

\begin{table*}[htbp]
\centering
\label{tab:table}
    \resizebox{\textwidth}{!}{
    \large
    \begin{tabular}{*{10}{c}}
        \toprule
       methods  & clear & jpeg,$\theta$=10 & jpeg,$\theta$=20 &  resize,$\theta$=0.1 & resize,$\theta$=0.2 &  saturation,$\theta$=10 &  saturation,$\theta$=20 &  Rotation,$\theta$=10 &  Rotation,$\theta$=20 \\
        \midrule
        Baluja(512)\cite{baluja2019hiding}  & 28.79 & 9.53 & 9.55 & 9.68 & 9.61 & 8.86& 8.46 &9.34 &9.16 \\
        HiNet(512)\cite{jing2021hinet}  & 33.93& 10.39& 10.52 & 10.67 & 10.69 & 15.44 & 14.18 &9.75 &9.93\\
        CRoSS(512)\cite{yu2024cross} & 25.30 & 16.11 & 16.45 & 17.42 & 17.50 & 12.61 & 10.67 &14.74 &13.59\\
        \midrule
        RoSMM(Gaussian) &20.35 & 19.81 & 20.30 & 13.60 & 20.1 & 19.89 & 18.22 &15.38 &14.03\\
        RoSMM(PCR) &16.62 & 15.45 & 15.75 & 12.93 & 15.81 & 14.97 & 12.62 &13.61 &13.21\\
        \bottomrule
    \end{tabular}
    }
    \caption{The PSNR (dB) performance results of different methods under various types of attacks. The higher the PSNR value, the better the image quality.}
    \label{tab:PRC_performance}
    \vspace{-1em}
\end{table*}

\begin{table}[htbp]
\centering
\begin{tabular}{c|c|cc|cc}
\toprule
\multirow{2}{*}{methods}
& \multirow{2}{*}{clean} & \multicolumn{2}{c}{Gaussian noise} & \multicolumn{2}{c}{Brightness}  \\
\cmidrule(lr){3-4} \cmidrule(lr){5-6} 
& & $\theta$=0.1 & $\theta$=0.2  & $\theta$=1 & $\theta$=2  \\
\midrule
Baluja  (256) & 0.934 &0.132 &0.069 &0.260 & 0.253 \\
Baluja  (512)   &0.920 &0.162 &0.074 &0.386 &0.375  \\
HiNet (256) &0.930    &0.014  &0.005    &0.576  &0.505   \\
HiNet (512) &0.933    &0.006  &0.005    &0.580  &0.545   \\
CRoSS (256)  &0.364    &0.171  &0.133    & 0.290 & 0.281   \\
CRoSS (512)  &0.853   &0.230 &0.117   & 0.578 & 0.546  \\
\midrule
RoSMM-w  &0.636    &0.222  &0.199    &0.586  &0.505  \\
RoSMM(ours)  &0.615    &\textbf{0.439}  &\textbf{0.366}  &\textbf{0.598}  &\textbf{0.581}  \\
\bottomrule
\end{tabular}
\caption{The SSIM performance results of different methods under various types of attacks. The closer the SSIM value is to 1, the better the image quality.}
\label{tab:ssim_performance}

\end{table}

In our proposed method, we also carefully designed the training loss functions. We observed that relying solely on reconstruction loss can result in overly smooth generated images, leading to a loss of crucial texture details. Therefore, we incorporated a perceptual loss function to better preserve the image’s fine details and textures. The perceptual loss leverages a pre-trained deep neural network (such as the VGG network) to extract feature representations of the image and compare the differences in these feature layers. The perceptual loss can be defined as:
\begin{equation}\label{eq:6}
L_{\text{perceptual}} = \left\| f_k(I_{\text{real}}) - f_k(I_{\text{fake}}) \right\|_1 \quad
\end{equation}
where \( I_{\text{real}} \) represents the real image, \( I_{\text{fake}} \) is the generated image, \( f_k \) denotes the feature extraction function at the \( k \)-th layer of the pre-trained deep network, and \( \left\| \cdot \right\|_1 \) indicates the L1 norm, used to compute the absolute difference between two feature maps.

The total loss function for our model combines perceptual loss, vector quantization loss, and commitment loss as a weighted sum, represented as:
\begin{equation}\label{eq:7}
 L_{\text{total}} = L_{\text{perceptual}} + \alpha L_{\text{embedding}} + \beta L_{\text{commitment}} 
\end{equation}
where \( \alpha \) and \( \beta \) are hyperparameters balancing the contributions of each loss term. With this loss function design, our model optimizes the quality and diversity of generated images while maintaining the discreteness of the encoder outputs, ensuring that the generated images retain high visual quality and rich texture details.

\subsection{More Robust Multi-Modal Watermarking Method}




We address the robustness of text watermarks under severe attacks and design a more resilient watermarking method. Text watermarks may be compromised during transmission, making it difficult to ensure 100\% accuracy, which directly impacts image watermarks, as shown in Figure \ref{fig:four}. This raises a key challenge: how to maintain image quality when text watermarks are damaged during transmission.

\begin{figure*}[htbp] 
\centering
\includegraphics[width=\textwidth]{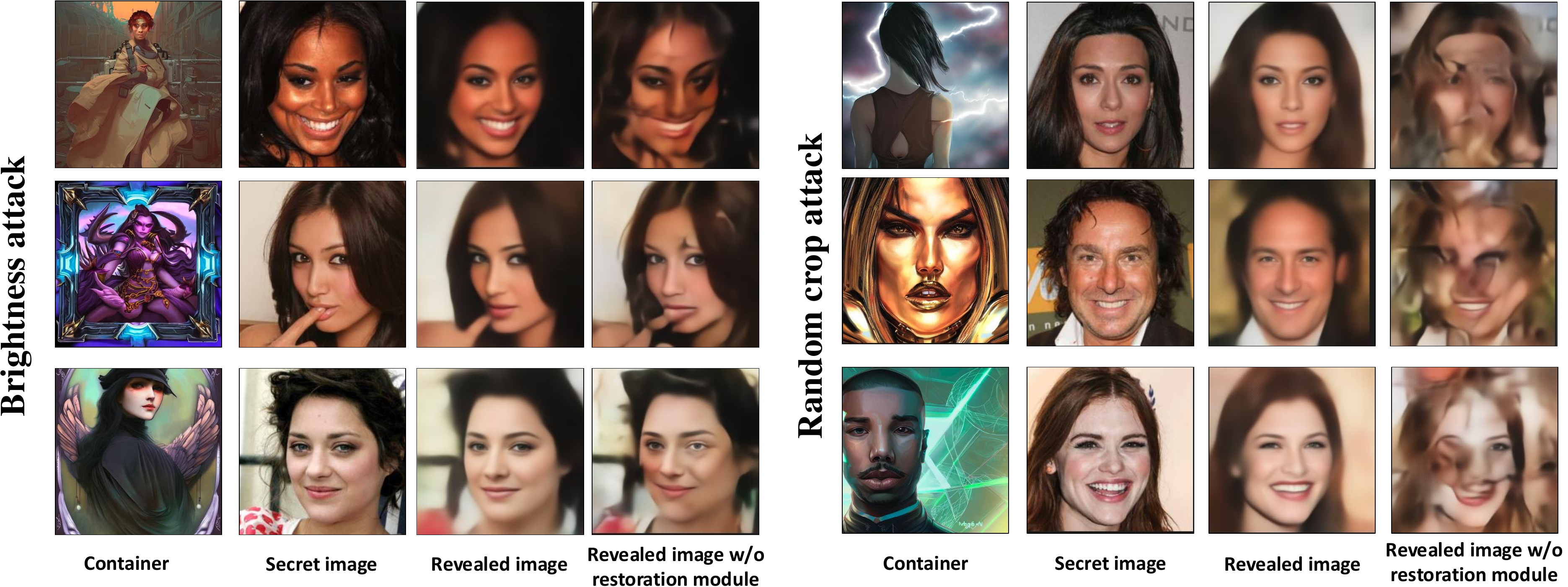} 
\caption{This figure compares the watermark imaging effects of the model under lossy transmission conditions with and without the image restoration module. The left column displays the watermark images without the restoration module, while the right column shows the imaging results after the introduction of the image restoration module.}
\label{fig:six}
\end{figure*}

To address this challenge,we introduce an image restoration module, as illustrated in Figure \ref{fig:five}. This module removes feature impurities through feature compression and restores clean features using upsampling techniques, thereby effectively preserving image quality. Specifically, the module compresses and upsamples feature maps to repair and enhance damaged features, ensuring visual quality even when watermark information is compromised. Figure  \ref{fig:six} compares the watermarking results with and without the restoration module, demonstrating its effectiveness in enhancing robustness and readability under damaged transmission conditions.

More specifically, The image restoration module employs an encode-decode structure with multiple downsampling and upsampling operations. The encoder consists of three downsampling blocks, each including a 3 $\times$ 3 convolutional layer, batch normalization, and ReLU activation, compressing features to form watermark encodings. The decoder, symmetric to the encoder, includes multiple upsampling blocks with 3 $\times$ 3 transposed convolutional layers, batch normalization, and ReLU activation. Skip connections are introduced in the initial downsampling stage to enhance high-frequency information recovery, ensuring clear watermark identification even under damaged transmission.

In summary, the proposed multi-modal watermarking framework is theoretically innovative and demonstrates excellent performance in practical applications. By integrating perceptual loss, an image restoration module, and multi-modal watermarking techniques, our framework effectively resists various attacks during watermark transmission, ensuring the robustness of the watermark while maintaining the integrity of the image watermark content restored from the text watermark.

\section{Experiment}
\subsection{Implementation Details}

\textbf{Experimental Settings } In this paper, we selected Stable Diffusion v2.1 \cite{rombach2022high}  as the conditional diffusion model, and implemented the Gaussian Shading model \cite{yang2024gaussian} for text watermarking within our multi-modal watermarking method. According to the specifications of the Gaussian Shading model \cite{yang2024gaussian}, we configured the hyperparameters as follows: \( fh = 2 \), \( fc = 2 \), and \( l = 1 \) and setting the binary watermark capacity at 2048 bits. To further validate the broad applicability of our approach, we introduced the xx method as an additional baseline model. Given the limitations of the PRC \cite{gunn2025undetectable} method in terms of watermark length, we made necessary adjustments to it, which, although compromising the model's robustness to some extent, ensured that sufficiently long watermark information could be processed by the model. Consequently, RoSMM (PCR) exhibits certain fluctuations in performance, whereas RoSMM(Gaussian) maintains consistent image generation quality with the employed vqvae model in an interference-free environment.

For seamless text-to-image transformation, we defined the adjacency matrix size as 16×16, and set the embedding depth of the codebook to 512 with a length of 256, allowing each codeword to be converted into an 8-bit binary encoding for efficient embedding in the text watermarking model, and the RoSMM-w means the RoSMM model without feature restoration modual. All experiments were conducted on an A6000 GPU, and our approach required no additional training or fine-tuning of the diffusion model, only the image-to-text and text-to-image transformation modules were trained. In addition, we have not identified any method that simultaneously achieves image steganography and text content watermarking. For comparison, we selected HiNet \cite{jing2021hinet}, Baluja \cite{baluja2019hiding} as a classic traditional image steganography method, and compared it with the latest diffusion-based image steganography method CRoSS \cite{yu2024cross}, demonstrating the robustness and readability of this method in image generation quality.

\textbf{Data Preparation} 
Regarding datasets,  we selected CelebA-HQ \cite{karras2017progressive} as our benchmark dataset. This dataset was generated by training a high-resolution Generative Adversarial Network (GAN) based on the original CelebA dataset \cite{liu2015faceattributes}. It contains a total of 30,000 images, each available in multiple resolutions, ranging from 64x64 to 1024x1024 pixels. Due to its high quality and diversity, CelebA-HQ is widely used across various computer vision domains, including object detection, image generation, and facial recognition. In this experiment, we randomly selected 100 images as our test set, with the remaining data used for training.

\begin{table*}[htbp]
\centering
\begin{tabular}{c|c|ccc|ccc|ccc}
\toprule
\multirow{2}{*}{methods} 
 & \multirow{2}{*}{clean}  & \multicolumn{3}{c}{Grussian noise} & \multicolumn{3}{c}{brightness}  & \multicolumn{3}{c}{Random Crop} \\
\cmidrule(lr){3-5} \cmidrule(lr){6-8}  \cmidrule(lr){9-11}
 & & $\theta$=0.05 & $\theta$=0.1 & $\theta$=0.2 & $\theta$=0.5 & $\theta$=1 & $\theta$=2 &  $\theta$=0.2 & $\theta$=0.4 & $\theta$=0.8 \\
\midrule
Gaussian Shading \cite{yang2024gaussian} & 99.2 & 83.2 & 72.8 & 63.5 & \textbf{99.0} & \textbf{98.4} & \textbf{95.0} & 58.0 & 71.6 & 89.9 \\
RoSMM(ours) & 96.8 & \textbf{84.0} & \textbf{75.8} & \textbf{71.7} & 95.0 & 94.8 & 94.2  & \textbf{69.8} & \textbf{72.2} & \textbf{90.1} \\
\bottomrule
\end{tabular}
\caption{Accurate(\%) results of the impact of image watermark robustness on text watermark Performance.}
\label{tab:absolution}
\end{table*}

\subsection{Comparison to Baselines}




In the experimental section, we examine various attack scenarios to evaluate the robustness of our proposed method in real-world applications. Specifically, we subjected container images embedded with secret images to several types of attack, including Random Crop, Gaussian noise, and Brightness attack. The hyperparameters $\theta$ represent the varying intensities of the aforementioned attacks. Following each attack, we extracted the secret image from the container to assess the robustness of watermark transmission.

To reduce the capacity load on text watermark transmission, our method RoSMM generates watermark images with a resolution of  256 $\times$ 256. We used PSNR (Peak Signal-to-Noise Ratio) and SSIM (Structural Similarity Index) as metrics to evaluate the robustness of other algorithms in watermarking at different image scales, where PSNR is used to quantify the loss of image quality, and SSIM is used to measure the structural similarity of images. As shown in Table \ref{tab:performance} \ref{tab:PRC_performance} and Table \ref{tab:ssim_performance}, compared to other methods, our multi-modal watermarking architecture demonstrates excellent robustness under multiple attack conditions.

In most attack scenarios, our method achieves the best performance. Only when the brightness attack strength is 0.5, HiNet \cite{jing2021hinet} outperforms our method. One possible reason is that when $\theta$ = 0.5, the attack has less impact on the watermark, making it difficult for other methods to surpass HiNet \cite{jing2021hinet} in image quality. However, even under this brightness attack, our method showed only slight performance degradation, whereas HiNet’s \cite{jing2021hinet} PSNR dropped sharply from 33.37 db to 23.73 db, reducing image quality by approximately 30\%, and Baluja \cite{baluja2019hiding} PSNR dropped  from 28.91 db to 8.90 db. Meanwhile, CRoSS \cite{yu2024cross} exhibited less fluctuation in performance, highlighting the advantages of diffusion models in image steganography. However, under high-intensity attacks, its PSNR and SSIM metrics remain slightly lower than our method, and in the case of Gaussian noise attacks, when $\theta$ = 0.2, its SSIM value drops to 0.005, while ours remains around 0.366.

Unlike traditional image steganography methods, our proposed multi-modal method, RoSMM, leverages the advantages of text watermarking to demonstrate superior robustness in watermark transmission. This method effectively combines the strengths of image and text watermarking, enabling the model to successfully transmit image watermark information with rich content through text watermarks.

\subsection{Internal Robustness Analysis of Multi-Modal Watermarks}




\begin{figure*}[htbp] 
\centering
\includegraphics[width=\textwidth]{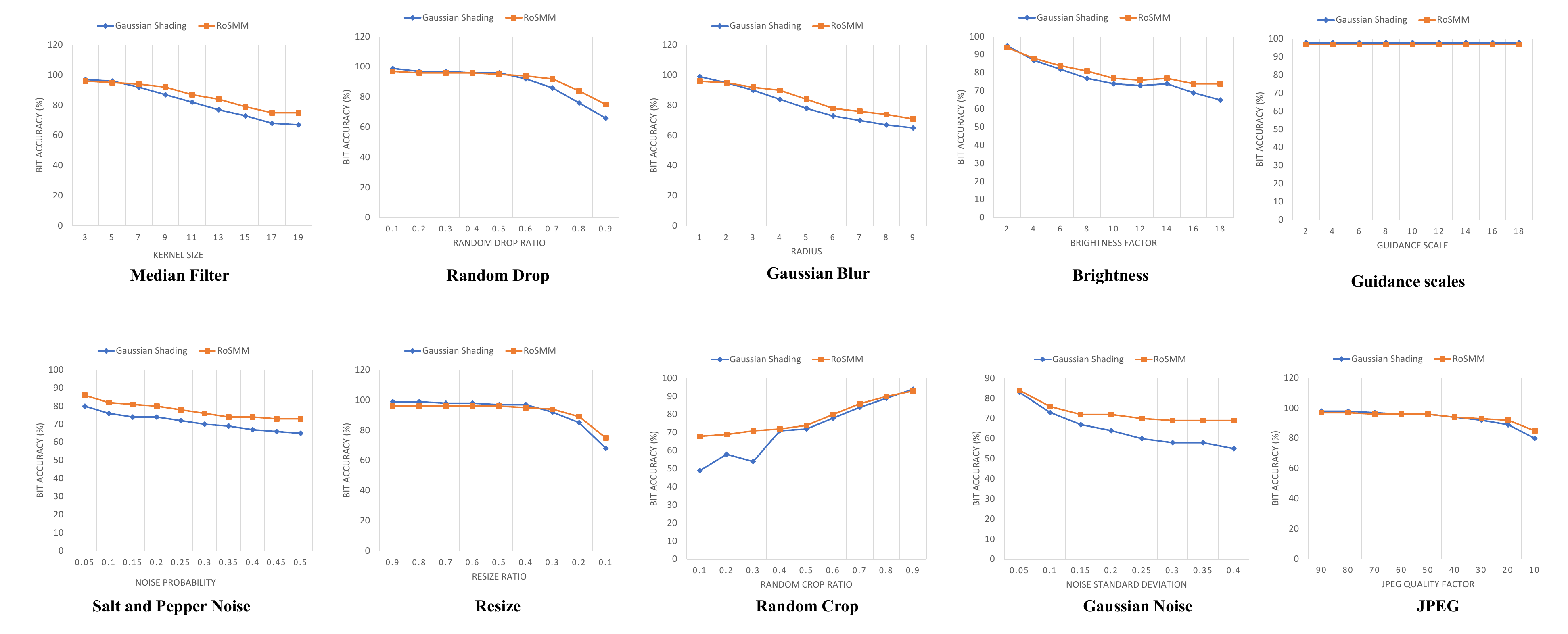} 
\caption{Perturbation Attack Results: We conducted ablation experiments on the proposed model with varying noise intensities}
\label{fig:ten}
\end{figure*}

In this section, we explore the impact of image watermarking and text watermarking technologies within the multi-modal watermarking framework and draw conclusions through experimental analysis. The framework does not simply concatenate image and text watermarks but instead creates a novel structure to integrate these technologies, fully leveraging their respective strengths and enabling mutual influence, thereby effectively harnessing the advantages of both text and image watermarking.

The Gaussian Shading model \cite{yang2024gaussian}, serving as the base for the text watermarking algorithm, offers strong robustness and can transmit over 512 bits of data. As shown in Table \ref{tab:absolution}, the Gaussian Shading model \cite{yang2024gaussian} maintains over 90\% accuracy under various attacks, highlighting its potential in real-world scenarios. However, its accuracy drops significantly under extreme conditions like Random Crop and Gaussian noise, falling to 58\% at $\theta$= 0.2 for Random Crop, indicating limitations in text watermarking.

To address this, we innovatively reframe text watermark repair as an image feature restoration problem. By improving image watermark quality, we indirectly enhance text watermark accuracy. As shown in Table \ref{tab:performance}, without the image restoration module, RoSMM-w exhibits lower image quality compared to RoSMM, dropping from 12.28 dB to 9.38 dB under Gaussian noise at $\theta$ = 0.2. This confirms the effectiveness of the image feature restoration module in enhancing watermark robustness. Additionally, Table \ref{tab:absolution} shows that improving image quality positively impacts watermark accuracy, especially when the initial accuracy is low. For example, under Random Crop at $\theta$ = 0.2, text watermark accuracy increases from 58.0\% to 69.8\% after image restoration. This improvement continues until the text watermark accuracy reaches 96\% or higher. As shown in Figure \ref{fig:ten}, we conducted ablation experiments by testing the proposed method with various attack intensities to evaluate whether the image restoration module can effectively improve the accuracy of text watermarking. The results demonstrate that, in the vast majority of cases under attack interference, the proposed method can adapt to almost all attack methods, and the text watermark accuracy of RoSMM consistently outperforms that of the baseline methods.

The proposed multi-modal watermarking content method achieves the synergy between image watermarking and text watermarking. Text watermarking ensures the robust transmission of image content, while image watermarking technology, in turn, provides the potential to enhance the transmission accuracy of text watermarks.



\section{Conclusion}

In this paper, we propose a multi-modal watermarking framework by integrating text content watermarking and image content watermarking into a unified system, introducing a novel multi-modal watermarking content method. This method leverages vector discretization in encoder-based vector quantization to achieve mutual transformation between text content watermarks and image content watermarks, addressing a key limitation in existing technologies. The inclusion of a restoration module effectively mitigates accuracy loss caused by various attacks, significantly enhancing the robustness of the watermark while maintaining the integrity of the image content. Experimental results demonstrate that the method exhibits strong resistance to almost all noise attacks, making it a secure and scalable solution for digital copyright protection.


{
    \small
    \bibliographystyle{ieeenat_fullname}
    \bibliography{main}
}
\end{document}